 \def\Title{\protect\textrm{%
 Translation of G.~L\"uders' \emph{\"Uber die Zustands\"anderung durch den Me\ss proze\ss
 }}}
 \def\arXiv{quant-ph/0403007v2}
 \def\Abstract{
 A translation and discussion of 
 G.~L\"uders, 
 \emph{Ann.~Phys. (Leipzig)} {\bf8} 322-328 (1951).
 }
\documentclass[aps,rmp,nobibnotes,nofootinbib,onecolumn,twoside,noshowpacs,draft]{revtex4}
 \usepackage[tbtags,sumlimits]{amsmath}
 \usepackage{amssymb}

 \newcommand{\Luders}{L\"uders}
 \newcommand{\Tr}[1]{{\rm Tr}\,(#1)}
 \newcommand{\KDelta}[2]{\delta_{#1\,#2}}
 \newcommand{\Zppkj}{Z_{kj}''}
 \newcommand{\Pt}{\widetilde{P}_j}
 \newcommand{\Prj}[1]{P_{#1}}
 \newcommand{\ket}[1]{\vert\,{#1}\,\rangle}
 \newcommand{\bra}[1]{\langle\,#1\,\vert}

\begin{document}
 \makeatletter
 \def\ps@titlepage{%
   \renewcommand{\@oddfoot}{}%
   \renewcommand{\@evenfoot}{}%
   \renewcommand{\@oddhead}{\emph{Ann.~Phys. (Leipzig)} {\bf 15} (2006)\hfill\arXiv}
   \renewcommand{\@evenhead}{}}
 \makeatother

\title%
 [Kirkpatrick -- Translation of L\"uders, \emph{``\"Uber die Zustands\"anderung \dots''}]%
 {\Title} 

\author{%
  \rm{K.~A.~Kirkpatrick}\\%
  \textit{New Mexico Highlands University, Las Vegas, New Mexico 87701}\\
  \footnotesize E-mail: \texttt{kirkpatrick@physics.nmhu.edu}}

\begin{abstract}
 \Abstract
\end{abstract}
 \maketitle
 \makeatletter\markboth{\hfill\@shorttitle\hfill}{\hfill\@shorttitle\hfill}\makeatother
 \pagestyle{myheadings}

\renewcommand{\thefootnote}{\fnsymbol{footnote}}
{\centering\textbf{\textsl{\large 
Concerning the state-change due to the measurement process%
}}
\\}\bigskip
\renewcommand{\thefootnote}{\arabic{footnote}}\setcounter{footnote}{0}

{\centering\textsl{By Gerhart L{\"u}ders}\\}\bigskip

{\centering\textbf{Abstract}\par}\medskip

The statistical transformation theory contains procedures for the computation of measurement probabilities, but requires for its completion a statement about the state-change due to the measurement process.  An ansatz suggested for this by J.~von Neumann is discussed and rejected. An ansatz is proposed for the state-change which is essentially identical to the ``reduction of the wave function.''  It permits a deepening of the concept of the compatibility of measurements.  Finally, measurements on constrained states are considered.  

{\centering\textbf{---------------}\\}


\bigskip{\centering\textbf{1.  Introduction}\par}\medskip

As is well known, the rules of the ``statistical transformation theory,'' along with knowledge of the ``state,'' allow the prediction of the statistical outcome of any measurement on an ensemble of identical and independent systems.  On the one hand, the state itself changes temporally in accordance with the Schr\"odinger equation, and on the other hand, it changes due to measurement processes made on the ensemble. Generally, the state of an ensemble can be known only by virtue of preceding measurement processes. To be complete and consistent, the statistical transformation theory must be supplemented with a statement regarding the change of a quantum-theoretical state due to the measurement process. 

It may not be redundant to emphasize that statements on the change of state due to measurement do not arise out of quantum theory itself through the inclusion of the measurement apparatus in the Schr\"odinger equation. Measurement, an act of cognizance, adds an element not already contained in the formulation of quantum theory.

The ansatz for the state-change due to measurement is obvious --- and generally recognized as correct --- in the case of the measurement of a quantity with simple eigenvalues (in the following, quantities and their associated operators are not differentiated). However, in the case of the measurement of degenerate quantities, we have, on the one hand --- at least for position measurements --- the statement of the ``reduction of the wave function,''%
\footnote{Cf., e.~g., W.~Pauli in \emph{Handbuch der Physik}, 2nd Ed., Vol.~{\bf24/1}, p.~83, Berlin 1933; esp.~Sec. 9. [English translation: W.~Pauli, \emph{General Principles of Quantum Mechanics}, Springer-Verlag, Berlin 1980, Sec.~9, esp.~p.~71.]} %
and, on the other hand, an ansatz due to J.~von Neumann.%
\footnote{J.~von Neumann, \emph{Mathematische Grundlagen der Quantenmechanik}, Berlin 1932.} %
However, the author cannot accept the latter's considerations, as will be discussed in detail.  The author emphasizes, though, that it was only through the study of the work of von Neumann that the definition of the problem became clear, and that the ansatz presented here for the change of a quantum-physical state due to the measurement process uses mathematical tools made available by von Neumann.  

The ansatz presented here is essentially identical to the ``reduction of the wave function''.  It permits two new conditions, verifiable at least in Gedanken-experiment, for the compatibility of measurements, and thereby a deepening of this concept from a physical point of view.  

1.  On an ensemble a quantity $R$ is measured, and that subensemble which gives a certain measured value $r_k$ is selected. On this subensemble, directly afterwards (so we can ignore any change of the state in accordance with the Schr\"odinger equation), the quantity $S$ is measured, and that subensemble which gives the measured value $s_j$ is selected. The measurements of the quantities $R$ and $S$ are to be called compatible with one another if a subsequent $R$-measurement is \emph{certain} to give the measured value $r_k$, regardless of the state of the ensemble before the first $R$-measurement. By execution of a sufficient number of measurements which are pairwise compatible in the sense defined here, a ``maximal observation'' in the sense of Dirac can be accomplished and a ``pure state'' manufactured and selected.  

2.  The second condition for the compatibility of measurements concerns measurements without a following selection of subensembles.  In this second sense the measurements of two quantities $R$ and $S$ are to be called compatible with one another if, by interposing an $R$-measurement (without subsequent selection), the outcome of the $S$-measurement is not affected.  These measurements can take place at different times.  

Using the proposed ansatz for the state-change due to measurement, we can determine the mathematical conditions the operators $R$ and $S$ must fulfill so that compatibility of the measurements is ensured in the senses 1 and 2.  We show that commutability of the operators $R$ and $S$ is necessary and sufficient for both senses of compatibility.  The mathematical proof is accomplished purely algebraically.  Analytic problems, such as questions regarding the domain of existence of the operators, are left unconsidered.  

Finally, we consider measurements of systems whose state is restricted by constraints (e.~g., systems of identical particles, and quantum electrodynamics in the formulation of Fermi).  It seems significant to the author that not only is the constraint satisfied under temporal change of the unobserved system in accordance with the Schr\"odinger equation, but that it is also satisfied after execution of a measurement process.  From this arises a restriction to ``allowed'' --- better called: physically possible --- measurements of such systems.

\bigskip{\centering\textbf{2. Ansatz for the state-change}\par}\medskip

Following von Neumann (loc.~cit.), our considerations will not be limited to ``pure states,'' which can be represented by a state vector (``wave function'') $\psi$, but will be extended to ``mixtures''.  If one includes measurements without a subsequent selection (see below), that is even absolutely necessary.  A unique positive and normalized%
\footnote{%
A hermitian operator is called ``positive'' if $(\psi,Z\,\psi) \geq 0$ for all $\psi$.  It is called ``normalized'' if $\Tr{ Z} = 1$ (Tr = trace) applies.
} %
hermitian operator $Z$ is assigned to the state of any ensemble consisting of identical and mutually independent systems.  

The operator assigned to the quantity $R$ possesses the spectral representation%
\footnote{%
All sums are to be extended over $k$, if the opposite is not expressly indicated.
} %
\[  R = \sum r_k P_k     \tag{1}\]
where the $r_k$ represent the eigenvalues (measured values) and the corresponding projection operators satisfy
\[  P_j\, P_k = \KDelta{j}{k}\, P_k,  \tag{2a}\]\vskip-1.5em
\[ \sum P_k = 1.  \tag{2b}\] 
(The treatment is simplified by assuming that all such operators possess pure point spectra).  Following von Neumann, the probability $w(r_k,Z)$ that the value $r_k$ is measured on the individual systems of the ensemble represented by the operator $Z$ is calculated as
\[  w(r_k,Z)=\Tr{P_k\,Z}.     \tag{3}\] 
This is thus determined solely by the state ($Z$) and the projection operator of the measured value $r_k$ (subspace of the Hilbert space).  The ansatz (3) --- at least so far --- is justified only partly through experiment, but mainly by its compelling simplicity.  

Regarding the state of the ensemble after the measurement, one may consider either the separate subensemble in which a certain measured value, e.g. $r_k$, occurs (\emph{measurement followed by selection of $r_k$}), or one may consider the complete ensemble again combined after the measurement process (\emph{measurement followed by aggregation}).  

In two special cases unanimity seems to exist regarding the way the state of an ensemble is changed by the measurement process:  

1.  If the eigenvalue $r_k$ is simple (i.~e., $\Tr{P_k} = 1$), then, in a measurement of $R$ followed by the selection of the $r_k$, $Z$ is transformed to 
\[ Z_k' = P_{\,[\Psi_k]} \cdot \Tr{P_{\,[\psi_k]}\,Z} \tag{4}\] 
with 
\[ R\,\psi_k=r_k\psi_k. \tag{5}\] 

The trace expression (3) gives the relative frequency with which $r_k$ is realized.  The projection operator $P_{\,[\psi_k]}$ projects any vector on the vector $\psi_k$.  If $Z$ represents a \emph{pure} state, then this can be represented by a state vector $\varphi$ (instead of by $Z$), and the ansatz (4) means that the selected subensemble can be represented after the measurement by the state vector $\psi_k$.  

2.  If all eigenvalues of $R$ are simple, then, in a measurement of $R$ followed by aggregation, $Z$ is transformed to the normalized mixture 
\[ Z' = \sum Z_k' = \sum P_{\,[\psi_k]}\, \Tr{P_{\,[\psi_k]}Z}. \tag{6}\] 

Regarding the state-change due to measurement of degenerate quantities, von Neumann suggested the following postulate:  Although the physical quantity $R$ does not itself distinguish a particular orthogonal system of eigenvectors, nevertheless this is done by the particular measuring apparatus.  To each individual apparatus which measures $R$, a completely determined orthogonal system $\psi_{kl}$ (running index $l$) therefore belongs to each particular $r_k$, using which (6) may then be applied with the corresponding summation over $k$ and $l$.

The ansatz of von Neumann directly gives a spectral representation of $Z'$.  However, two serious concerns may be raised against it:  

1.  The measurement of a highly degenerate quantity permits only relatively weak assertions regarding the considered ensemble.  For that reason, the thereby resulting change in state should likewise be small, in particular with a subsequent selection, whereas the ansatz of von Neumann yields a most complicated mixture.  The extreme case is provided by ``measurement'' of the unit operator.  Nothing is revealed about the system, which should survive the ``measurement process'' uninfluenced.%
\footnote{For the moment we accept the fiction that every hermitian operator corresponds to a measurable quantity, although from physical considerations we believe we must in principle reject this.}  

2.  One would expect, according to formula (3) for the computation of the measurement probabilities, that the state-change depends only on $R$ (and, of course, on $Z$), thus only on the projection operators $P_k$.  

The ansatz to be presented here results almost inevitably if the two concerns raised against von Neumann's ansatz are accepted as fully justified.  We therefore postulate 

1. In a measurement of $R$ followed by selection of $r_k$, $Z$ is transformed to 
\[      Z_k' = P_k Z P_k.   \tag{7}\] 
$Z_k'$ is not normalized (see note 3), but instead is chosen so that the trace shows the relative frequency of the occurrence of $r_k$ in the ensemble.  In particular, if $Z$ represents a pure case, then the ansatz (7) ensures that, \emph{after} a measurement followed by selection of $r_k$, a pure case is again present.  

2.  Again, if an aggregation of the entire ensemble takes place after the measurement of $R$, then, in consequence of (7), $Z$ is transformed to 
\[     Z' = \sum Z_k' =\sum P_k Z P_k    \tag{8}\] 
It can be proven by means of elementary algebraic calculations that, because the projection operators $P_k$ are hermitian, the operators $Z_k'$ and $Z'$ are positive and hermitian if $Z$ has these characteristics. Likewise, $Z'$ is normalized if $Z$ is.  

(4) and (6) arise, as can be shown, as special cases of (7) and (8), if the measured values ($r_k$ or all values, respectively) are simple.  In addition, the ansatz (7) shows exactly what is meant by the expression ``reduction of of the wave function''.

\bigskip{\centering\textbf{3.  Compatibility of measurements}\par}\medskip

First we examine more closely, under the use of the ansatz (7), the first condition for the compatibility of measurements given in the introduction.  After execution of an $R$-measurement and selection of $r_k$, a subsequent $S$-measurement and selection of $s_j$ transforms $Z$ to the (unnormalized) state 
\[   \Zppkj = \Pt P_k Z P_k \Pt .  \tag{9}\] 
The operator $P_k$ corresponds to the eigenvalue $r_k$ in the spectral representation of $R$ (see (1)) and the operator $\Pt$ corresponds to the eigenvalue $s_j$ in the spectral representation of $S$.  The statement that it is \emph{certain} that $r_k$ will be measured on the state $\Zppkj$ can be formulated
\[ \Tr{\Zppkj\,P_l} = 0\quad \text{for all } l\neq k.   \tag{10}\] 
The requirement that (10) should hold for all $Z$ and all $k$ proves to be equivalent to the commutability of the two families of projections
\[ \Pt P_l = P_l \Pt   \tag{11}\] 
and this is well-known to be equivalent to the commutability of the operators $R$ and $S$. 

The compatibility of the measurements in this first sense, (10), follows immediately from these projector commutation relations by means of (11) and (2a). The converse ((11) as a consequence of (10)) is somewhat more difficult to prove.

Since (10) is to apply for all $Z$, in particular then for all P[$\varphi$], it follows first that 
\[  P_k\Pt P_l \Pt P_k = 0.  \tag{12}\] 
One may conclude, using the following lemma, that
\[  P_l \Pt P_k = 0.  \tag{13}\] 
Lemma:  If B is positive and hermitian, and if $C^{*}BC=0$ (* = hermitian adjoint operator), then BC=0.  

(Proof by application of the Cauchy-Schwarz inequality to the ``scalar product'' $(\psi,\,\varphi)_B\equiv(\psi,\,B\,\varphi)$.  The lemma is applied by setting $B = P_l$ and $C=\Pt P_k$.)  

Summing (13) over all $k\neq l$, it follows by (2b) that 
\[  P_l\Pt = P_l\Pt P_l;   \tag{14}\] 
the hermitian adjoint of this is
\[  \Pt P_l = P_l\Pt P_l . \tag{15}\] 
From (14) and (15) together we obtain (11), as was to be proven.  

The \emph{second condition} for the compatibility of measurements is that the outcome of a measurement of the quantity $S$ is not changed by an intervening $R$-measurement.  This statement may be formulated mathematically, using (3) and (8), as follows:  
\[  \Tr{\Pt\cdot\sum P_k Z P_k } = \Tr{\Pt Z}\quad \text{for all } j . \tag{16}\] 
The left side gives the probability to measure $s_j$ after a preceding $R$-measurement and the right side yields the corresponding probability without the intervention of an $R$-measurement.  

The validity of (16) for all $Z$ follows from (11) and thus is equivalent with the commutability of $R$ with $S$.  Again it needs to be proven that (11) is a consequence of (16), while the converse is seen directly.  

Exactly as (12) was obtained from (10), it follows from (16) that
\[  \sum P_k \Pt P_k = \Pt .  \tag{17}\] 
By multiplication with $P_l$ on the left, (14) follows immediately because of (2a).  Thus the statement is proven.  

Both conditions for compatibility, despite the outward appearance of the definitions, have been proven as symmetric in $R$ and $S$ and as essentially equivalent.  The second condition is, however, somewhat more general than the first, since it can be expanded to measurements at different times.  

In the Heisenberg representation one keeps $Z$ temporally constant and rolls over the temporal change onto the operators.  In generalization of the second condition, it is the case that, for arbitrary state $Z$, an $R$-measurement at the time $t_1$ does not change the outcome of an $S$-measurement at the (later) time $t_2$ if and only if the operators $R(t_1)$ and $S(t_2)$ commute with one another (in the Heisenberg representation).  This statement is occasionally used in the literature.  It becomes a provable statement, however, only when a definite ansatz for the state-change due to measurement has been constructed.

\bigskip{\centering\textbf{4.  Systems with constraints}\par}\medskip

Sometimes in applications the demand arises that the wave functions $\psi$ (or general states $Z$, cf.~(20)) fulfill a constraint 
\[  N \psi = 0.  \tag{18}\] 

Two examples shall clarify the meaning.  

1.  In systems which consist of several identical particles, a symmetry condition is to be satisfied which can be brought into the form (18).  

2.  With quantum electrodynamics in the formulation of Fermi%
\footnote{Cf.~G.~Wentzel, \emph{Einf\"uhrung in die Quantentheorie der Wellenfelder}, Vienna 1943;  esp.~Chap.~IV.} %
the wave functional must satisfy the set of constraints
\[ \left( \sum_{\mu=1}^{\psi}
  \frac{\partial A_{\mu}(x)}{\partial x_{\mu}}\right)\psi=0. \tag{19}\] 

Considering a state $Z$,
\[ NZ=0   \tag{20}\] 
is equivalent to the fact that $Z$ can be composed only of such pure states which fulfill (18).  

Usually it is shown only that, from the validity of (18) and/or (20) for a certain time, such validity follows for all later times if the ensemble is left undisturbed.  That is certainly the case if the Hamilton operator $H$ commutes with $N$.  It seems, however, to be just as important also to prove that, upon measurement of a quantity $R$, the state-change respects the validity of (18) and/or (20).  

One confirms easily, as a consequence of the ansatz (7), that (20) is also fulfilled for $Z_k'$ if $R$ (i.~e., the projection operator $P_k$) commutes with $N$ (resp.~the set of constraints).  For then it follows that
\[ NZ_k'\equiv N P_k Z P_k = P_k N Z P_k = 0.   \tag{21}\] 

On this basis we propose the following \emph{measurability postulate}:  If all admissible states of an ensemble of physical systems are restricted by a constraint (20) (or by a set of mutually compatible constraints), then it is physically possible to measure only such quantities whose associated operators commute with the constraint $N$ (or with the set of constraints).  

If one applies this postulate to the two examples mentioned at the beginning of the section, then one recognizes that the measurement of quantities which are symmetrical in the particles is not forbidden by the postulate and that likewise the measurements of the electrical and magnetic field strengths do not violate the postulate, whereas that of the scalar potential does.  

\bigskip
I owe essential viewpoints to numerous discussions with Dr.~H.~Fack (Hamburg) regarding fundamental questions of the quantum theory. I gladly and cordially thank Dr.~Fack.

\bigskip
{\centering Hamburg, Institut f\"ur Theoretische Physik der Universit\"at.\\}  
\smallskip{\centering(Received at editorial office on 18 October 1950.)\\}


\bigskip\bigskip\noindent{\small\textbf{Translator's note.} Immediately following (7), $Z_k'$ replaces the misprint $Z_k$; no other substantive changes were made.}

\bigskip\bigskip
\noindent\textbf{Discussion}\medskip  

In this 1951 paper \Luders\ introduced an ansatz for the measurement transformation of the quantum state which differed in the case of degeneracy from the von Neumann ansatz of 1932 \cite{vonNeumann55tr}.  He  ``deepened the concept of compatibility,'' introducing two probability expressions for the compatibility of a pair of observables and showing that, under his proposed ansatz, each condition is equivalent to the commutativity of the associated operators. And he proposed that an observable is measurable in the presence of a null constraint only if its operator commutes with the constraint (supported the following year by Wigner \cite{Wigner52} and proven generally by Araki and Yanase nine years later \cite{ArakiYanase60}). 

Let us examine the relation between von Neumann's  and \Luders' ansatz. We carry this out in terms of the measurement of an observable $R$ whose associated operator (cf.~\Luders' Eq.~(1)) has eigenstates $\{\psi_j : j\in D\}$ and has the spectral representation $R=\sum_{k\in E}r_k\Prj{k}$, with $\Prj{k}=\sum_{j\in D_k}\Prj{[\psi_j]}$, $\bigcup_{k\in E} D_k=D$, and $D_k\cap D_{k'}=\varnothing, k\neq k'$. (Non-singleton $D_k$'s represent degeneracy of the eigenvalues.)

Von Neumann \cite[p.~348]{vonNeumann55tr} approached the measurement of a degenerate observable through the claim that the measurement of an observable is, at the same time, a measurement of any function of that observable. It follows that a measurement of the degenerate observable $R$ may be accomplished by a measurement of the non-degenerate (distinct $\{q_j\}$) observable $Q=\sum_{j\in D}q_j\Prj{[\psi_j]}$, thus that a measurement of $R$ must transform the statistical operator as does a measurement of $Q$ --- leading to von Neumann's ansatz, $Z'=\sum_{s\in D}\Prj{[\psi_s]}\Tr{Z\Prj{[\psi_s]}}= \sum_{s\in D}\Prj{[\psi_s]}Z\Prj{[\psi_s]}$. Von Neumann noted that this expression is ``unambiguous'' only for $Z$ such that $\Prj{k}Z\Prj{k}\varpropto\Prj{k}$ --- otherwise, for a degenerate eigenvalue, it represents a particular choice of eigenstates within the degeneracy subspace.

\Luders\ raised two concerns regarding the von Neumann ansatz. The first concern required that if $R$ is the (fully degenerate) identity then it should be that $Z'=Z$, the second required that the transformation should depend only on the $\{\Prj{k}\}$. Together, these led \Luders\ ``almost inevitably'' to the ansatz  
$Z'=\sum_{k\in E}\Prj{k}Z\Prj{k}= \sum_{k\in E}\sum_{s,t\in D_k}\Prj{[\psi_s]} Z \Prj{[\psi_t]}$. By leaving the degenerate eigenspaces invariant, this avoids the ambiguity noted by von Neumann.  

\Luders\ places von Neumann's ansatz in opposition to the ``reduction of the wave function,'' and says that his own ansatz ``is essentially identical to'' and ``shows exactly what is meant by the expression'' ``reduction of the wave function''. Clearly, that expression is quite important to \Luders, and can only mean the invariance of eigenstates under \Luders' ansatz, lacking in von Neumann's.

\Luders\ considered von Neumann's ansatz for measurement involving a degenerate eigenvalue to be fundamentally incorrect; he raised ``serious concerns'' about it, he ``could not accept'' it, he ``rejected'' it. These phrases, however, misstate the issue, which cannot be the ``truth'' of one ansatz or the other, but rather the question of which (if either!) correctly represents the particulars of the measurement situation. Let us consider several versions of those particulars.

On the one hand, if the interaction between the system ($\mathcal{S}$, say) and the measuring apparatus ($\mathcal{M}$, say, initialized in the state $\xi_0$, the set of states $\{a_j\}$ orthonormal) resulted in the transformation $\psi_j\otimes \xi_0\rightarrow\psi_j\otimes a_j, \forall j\in D$, then in $\mathcal{S}$ the transformation would be described by von Neumann's ansatz. If, on the other hand, that measurement interaction were to result in the transformation $\psi_j\otimes \xi_0\rightarrow\psi_j\otimes a_k, \forall j\in D_k,\forall k\in E$, then in $\mathcal{S}$ the resulting transformation would be described by \Luders' ansatz. Von Neumann's ansatz represents a degeneracy-breaking measurement, while \Luders' ansatz represents a repeatable measurement of degenerate eigenvalues which, in addition, leaves eigenstates undisturbed. (This analysis shows that each is correctly called an ``ansatz,'' for each is a shorthand rule for a result properly obtained through lengthier physical calculation.)

We might take a more ``physical'' look at the measurement situation in terms of the issue of distinguishability. The von Neumann measurement interaction makes a particular selection of eigenstates within each degeneracy subspace, fully distinguishing each eigenstate with \emph{welcher Weg} information (the orthonormal $\mathcal{M}$-states $\{a_j\}$). This full distinguishability prevents coherence among the eigenstates, resulting in a mixture and ``breaking'' the degeneracy. The \Luders\ measurement interaction, on the other hand, correlates all the $\{\psi_j : j\in D_k\}$ belonging to the eigenvalue $r_k$ with a single distinct $\mathcal{M}$-state $a_k$; this distinguishes the eigenstates belonging to each eigenvalue from those belonging to any other eigenvalue, but all those belonging (degenerately) to a single eigenvalue are left indistinguishable, lacking any \emph{welcher Weg} information. This indistinguishability maintains coherence among the $\{\psi_j : j\in D_k\}$ and thus leaves their resultant state --- an eigenstate of the degenerate eigenvalue --- undisturbed.

\Luders\ stated his two ``concerns'' as Hilbert space expressions; let us restate their implications in physical terms. The requirement arising from \Luders' first concern implies (much in the spirit of the compatibility conditions) that interposing a \Luders\ (degeneracy-respecting) measurement of $R$ between two successive von Neumann (degeneracy-breaking) measurements of $R$ does not disturb the repeatability of the von Neumann measurements, and this holds over the range of ``ambiguity'' of the von Neumann measurements. The requirement arising from \Luders' second concern implies that the measurement of a degenerate eigenvalue must be repeatable (in parallel with von Neumann's requirement that the measurement of a \emph{non}-degenerate eigenvalue be repeatable). Both in the formal terms \Luders\ used, and in these more physical terms, the requirement arising from the second concern seems the more compelling. 

But \Luders\ might have considered (we avoid anachronism, using only concepts well-established prior to 1951) a more general (i.~e., not necessarily repeatable) degenerate $R$-measurement interaction resulting in the measurement transformation  
$\psi_j\otimes\xi_0\rightarrow\theta_j\otimes a_k, \forall {j\in D_k}, \forall k\in E$. This measurement transformation is represented by the ansatz $Z'=\sum_{k\in E}\Theta_k Z {\Theta_k}^{\!*}$, with 
$\Theta_k=\sum_{s\in D_k}\ket{\theta_s}\bra{\psi_s}$; because the interaction is unitary, the $\{\theta_s:s\in D_k\}$ are orthonormal. The restriction to eigenvalue-repeatable measurement (\Luders' second concern) requires that each $\theta_j$ lie within the degeneracy subspace of its corresponding $\psi_j$; it follows (for $D_k$ of finite cardinality) that $\{\theta_s:s\in D_k\}$ is an orthonormal basis of the $k$-th degeneracy subspace, that ${\Theta_{k}}^{\!*}\Theta_{k'}=\Theta_{k'}{\Theta_{k}}^{\!*}=\delta_{kk'}P_k$, and that 
$\Theta_k P_{k'}=\delta_{kk'}\Theta_k$. (Though the $\{\theta_j:j\in D_k\}$ are coherent, because this ansatz does not satisfy \Luders' first concern, it does not leave eigenstates invariant.) 

Under this generalized ansatz, both of \Luders' conditions of compatibility remain equivalent to commutativity: 
Applying the general $R$ and $S$ measurements (both eigenvalue-repeatable) to the first compatibility condition, \Luders' (9) becomes $\Zppkj = \Phi_j\Theta_k Z {\Theta_k}^{\!*}{\Phi_j}^{\!*}$ ($\Phi_j$ is the $S$-measurement transformation operator). Following \Luders' argument, applying $\Theta_{k'}{\Theta_{k}}^{\!*}=\delta_{kk'}P_k$, (13) then becomes $P_l\Phi_j P_k=0, l\neq k$. Comparing the results of summing this over $l$ and of summing this over $k$, we see that $P_k\Phi_j=\Phi_j P_k$. Multiplying this result on the left by ${\Phi_j}^{\!*}$, we have ${\Phi_j}^{\!*}P_k\Phi_j=\Pt P_k$; being hermitian, this gives (11), as was to be shown. 
Next, applying the general $R$-measurement ansatz to the second compatibility condition, \Luders' (17) becomes $\sum_k{\Theta_k}^{\!*}\Pt\Theta_k=\Pt$. Multiplying this on the right and on the left, respectively, by $P_{k'}$ and using $\Theta_k P_{k'}=\delta_{kk'}\Theta_k$ and its conjugate, respectively, yields (11). 

Thus, requiring of the measurement interaction only eigenvalue repeatability suffices%
\footnote{%
It suffices for the assumed finite degrees of degeneracy. P.~Busch (personal communication) has provided a nice counterexample for the infinite case.
} %
to establish the equivalence of compatibility with commutability, even though such measurement does not necessarily satisfy the ``reduction of the wave function'' with eigenvector invariance.


\bigskip\bigskip\noindent{\small\textbf{Acknowledgement} 
This translation%
\footnote{%
Original at: 
{\footnotesize\texttt{www.physik.uni-augsburg.de/annalen/history/historic-papers/1951\_443\_322-328.pdf}}
} %
and accompanying discussion have benefitted greatly from the generous efforts of Paul Busch, to whom I am most grateful. Of course, I bear full responsibility for any remaining awkwardness or error. 
}



\begin{thebibliography}{1}
%
\bibitem[Araki and Yanase(1960)]{ArakiYanase60}
Araki, H. and Yanase, M. (1960).
``Measurement of quantum mechanical operators,'' \emph{Phys. Rev.} {\bf 120},
  622--626.

\bibitem[{v}on Neumann(1955)]{vonNeumann55tr}
{v}on Neumann, J. (1955).
{\em Mathematical Foundations of Quantum Mechanics},
Princeton University Press, Princeton, N. J.,
\ R. T. Beyer, translator. \ Originally published as \emph{Mathematische Grundlagen der
  Quantenmechanik}, Springer, Berlin, 1932.

\bibitem[Wigner(1952)]{Wigner52}
Wigner, E.~P. (1952).
``Die {M}essung {q}uantenmechanischer {O}peratoren,'' \emph{Z. Phys.} {\bf
  133}, 101--108.
%
\end{thebibliography}
\end{document}